\documentclass[sn-mathphys,Numbered]{sn-jnl}% Math and Physical Sciences Reference Style
\usepackage{graphicx}
\usepackage{amssymb,amsmath}
\usepackage{gensymb}
\usepackage[Symbolsmallscale]{upgreek}

\def\unit#1{\mathord{\thinspace\rm #1}}
\DeclareMathAlphabet\mathbfcal{OMS}{cmsy}{b}{n}
\newcommand{\sgn}{\text{sgn}}
\renewcommand\figurename{Figure}

\begin{document}

\title[ ]{Probing miniband structure and Hofstadter butterfly in gated graphene superlattices via magnetotransport}


\author*[1,2]{\fnm{Alina} \sur{Mre\'nca-Kolasi\'nska}}\email{alina.mrenca@fis.agh.edu.pl}

\author*[2,3]{\fnm{Szu-Chao} \sur{Chen}}\email{scchen@nfu.edu.tw}

\author*[2]{\fnm{Ming-Hao} \sur{Liu}}\email{minghao.liu@phys.ncku.edu.tw}

\affil[1]{\orgname{AGH University}, \orgdiv{Faculty of Physics and Applied Computer Science}, \orgaddress{\street{al. Mickiewicza 30}, \postcode{30-059} \city{Krak\'ow}, \country{Poland}}}

\affil[2]{\orgdiv{Department of Physics}, \orgname{National Cheng Kung University}, \orgaddress{\city{Tainan} \postcode{70101}, \country{Taiwan}}}

\affil[3]{\orgdiv{Department of Electro-Optical Engineering}, \orgname{National Formosa University}, \orgaddress{\city{Yunlin}, \country{Taiwan}}}

%%==================================%%
%%			 	abstract 			%%
%%==================================%%

\abstract{The presence of periodic modulation in graphene leads to a reconstruction of the band structure and formation of minibands. In an external uniform magnetic field, 
a fractal energy spectrum called Hofstadter butterfly is formed. Particularly interesting in this regard are superlattices with tunable modulation strength, such as electrostatically induced ones in graphene. 
We perform quantum transport modeling in gate-induced square two-dimensional superlattice in graphene and investigate the relation to the details of the band structure.
At low magnetic field the dynamics of carriers reflects the semi-classical orbits which depend on the mini band structure. We theoretically model transverse magnetic focusing, a ballistic transport technique by means of which we investigate the minibands, their extent and carrier type. We find a good agreement between the focusing spectra and the mini band structures obtained from the continuum model, proving usefulness of this technique. %positions of van Hove singularities 
At high magnetic field the calculated four-probe resistance fit the Hofstadter butterfly spectrum obtained for our superlattice.
Our quantum transport modeling provides an insight into the mini band structures, and can be applied to other superlattice geometries.}

%\keywords{keyword1, Keyword2, Keyword3, Keyword4}



\maketitle

\section*{Introduction}

Graphene, a 2D material characterized by a linear low-energy dispersion relation, hosts charge carriers named Dirac fermions due to the resemblance of relativistic (massless) particles described by the Dirac equation.
Modifying the underlying graphene lattice by a smooth periodic potential can affect the band structure %in even more extraordinary ways, 
through folding of the pristine graphene Dirac cone into mini bands %
\cite{Wallbank2013}, formation of the secondary Dirac points, and anisotropic renormalization of velocity \cite{Park2008, Park2008prb, Brey2009, Barbier2010, Kang2020}. 
Periodic modulation has been obtained in graphene through chemical functionalization \cite{Sun2011}, placing graphene on self-assembled nanostructures \cite{Zhang2018}, and by stacking graphene together with aligned or slightly misoriented hexagonal boron nitride (hBN), resulting in periodic moir\'e modulation which generates hexagonal superlattices (SLs) \cite{Xue2011, Decker2011, Yankowitz2012, Ponomarenko2013, Hunt2013, Dean2013, Yu2014}.
Moir\'e SLs were also created in low-angle twisted graphene bilayers
, followed by van der Waals structures made up of few layers of graphene \cite{Burg2019, Shen2020, Liu2020, Lin2020, deVries2020, Rickhaus2021} and other 2D materials \cite{WangLujun2019, WangZihao2019}. %including transition metal dichalcogenides \cite{Jin2019, Seyler2019, Tran2019, Alexeev2019, Merkl2019}. 
%The progress in the development of SLs allowed observation of secondary Dirac points \cite{Yankowitz2012, Ponomarenko2013}, as well as Fabry-P\'erot interference \cite{Handschin2017, Kraft2020} and magnetic focusing of fermions residing in the moir\'e minibands \cite{Lee2016, Berdyugin2020}. 
%SLs are also suitable for the observation of the self-similar energy spectrum called
SLs are suitable for the observation of the self-similar energy spectrum called
Hofstadter butterfly \cite{Dean2013, Hunt2013} and Brown-Zak oscillations \cite{Kumar2017, Barrier2020} that occur when the magnetic flux through the superlattice unit cell is of the order of the magnetic flux quantum, $\phi_0=h/e$, and in pristine 2D crystals require unattainable magnetic fields. Also worth mentioning are the %collective 
many-body phenomena present in moir\'e SLs \cite{Wang2015, Andrews2020}.

\begin{figure}[b]
\includegraphics[width=\columnwidth]{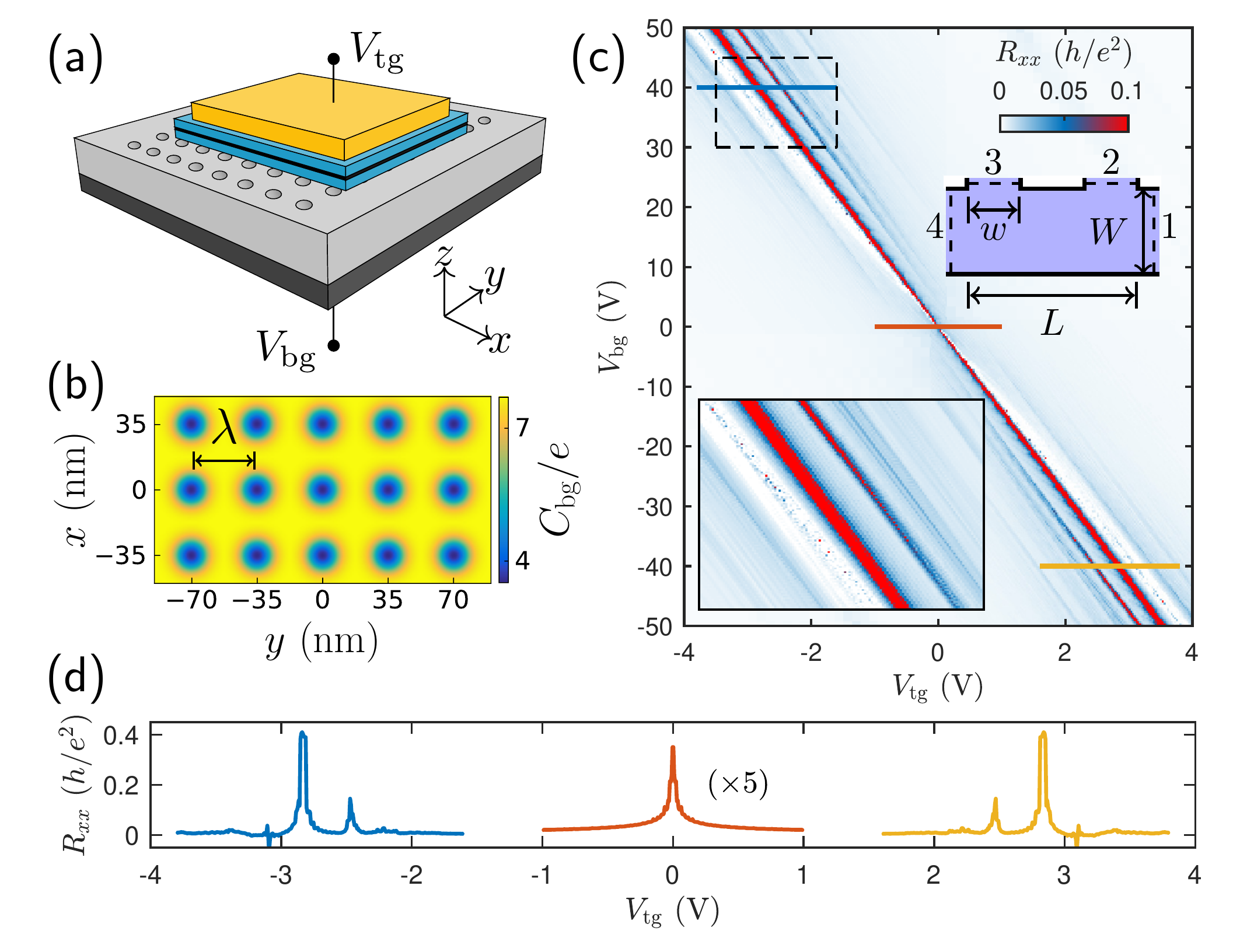} 
\caption{
\textbf{Zero magnetic field characterization of the superlattice.}
(a) Sketch of the gated superlattice system. The backgate (dark gray) lays below the patterned substrate (light gray), which induces modulated potential in graphene sandwiched between two hBN layers (blue). The global top gate is shown in yellow. (b) The backgate capacitance profile with the periodicity $\lambda=35$ nm in the units of $10^{12}\ \mathrm{cm}^{-2}\mathrm{V}^{-1}$. (c) Longitudinal resistance $R_{xx}$ as a function of the backgate and top gate voltages in a four-terminal device shown in the right inset. Left inset: zoom of $R_{xx}$ in the boxed region. (d) $R_{xx}$ line cuts as marked with the respective line colors in (c).} 
\label{fig:fig1}
\end{figure}
Despite their potential, 
artificial lattices tailored by chemical methods or moir\'e SLs suffer from inability to tune the strength of the periodic potential.
In moir\'e SLs the period can be tuned to some extent via the rotation angle between the stacked layers, but they are inherent of a hexagonal symmetry. 
Precise control over the SL geometry, period, and strength is vital for the band structure engineering. The above limitations can be circumvented in
electrostatic gate-induced SLs that allow an arbitrary design via the gates geometry, with the gate voltage being a knob for the potential strength. The experimental attempts to create gated SLs in graphene included 1D arrays of metal gates \cite{Dubey2013, 
Drienovsky2014, Kuiri2018}, followed by patterned dielectric substrates \cite{Forsythe2018, Li2021}, and few-layer graphene patterned bottom gates for 1D \cite{Drienovsky2018, Ruiz2022} and 2D SLs \cite{Huber2020, Huber2022} with down to sub-20 nm periods \cite{Ruiz2022} manifesting the flexibility of this approach. 
With the recent advance in the fabrication techniques, gated SLs with the period of a few tens of nanometers are achievable with good device quality and long electron mean free path \cite{Li2021}. This enabled observation of commensurability oscillations \cite{Drienovsky2018, Li2021, Huber2022}, Hofstadter butterfly \cite{Huber2020, Ruiz2022} and Brown-Zak oscillations \cite{Huber2022} at magnetic fields of the order of a few tesla.
This is more affordable compared to about 25 T required for graphene/hBN SL, where, due to small lattice constant mismatch of 1.8\%, the SL period can reach up to 14 nm for aligned lattices.

While transport in quantizing magnetic field in gated SLs has been thoroughly studied, the intermediate magnetic field regime remained mostly unexplored. %(however, note Ref.\ \cite{Kang2020} for 1D SL). 
In the semiclassical treatment, at low magnetic field fermions undergo cyclotron motion that can be probed in transport measurements via transverse magnetic focusing (TMF). 
This technique has been used to experimentally probe the band structure in pristine mono-, bi-, and tri-layer graphene \cite{Taychatanapat2013}, graphene/WSe$_2$ heterostructures \cite{Rao2023}, and moir\'e superlattices \cite{Lee2016, Berdyugin2020}, and theoretically considered in graphene $pn$ junctions \cite{Milovanovic2014, Chen2016} and 1D gated SLs \cite{Kang2020}. 

In this work, we perform a theoretical study of the TMF in 2D gate-induced square SLs, and analyze the relation between the observed TMF spectra and the miniband structure. 
This is complemented by the investigation of magnetotransport in the quantum Hall regime, where we observe signatures of Hofstadter butterfly, matching the numerical Hofstadter spectrum calculated for our gated SL. 
Our study is performed using quantum transport calculations for multiterminal structures, considering realistic experimental conditions.
To model a realistic geometry, we consider SL induced by a patterned dielectric substrate \cite{Forsythe2018, Li2021} with a uniform global backgate underneath the dielectric layer, and the graphene sheet sandwiched between two hBN layers lying on top [\autoref{fig:fig1}(a)]. The hBN/graphene/hBN sandwich is covered by a global top gate. The voltage applied to the back gate $V_{\mathrm{bg}}$ controls the strength of the periodic modulation, while the top gate voltage $V_{\mathrm{tg}}$ is used to tune the carrier density across the SL. We follow the design of Ref.\ \cite{Forsythe2018}, with a square lattice etched in SiO$_2$ substrate, with a lattice period $\lambda=35$ nm [\autoref{fig:fig1}(b)]. For the ease of the calculation, we use a model function $C_{\mathrm{bg}}(x,y)$ [see \autoref{fig:fig1}(b)] that approximates the electrostatically simulated capacitance, obtained previously by some of us \cite{Chen2020}. Previous studies \cite{Kraft2020, Huber2020} and modeling of previous experiment on magnetic focusing in monolayer graphene \cite{Taychatanapat2013} show good agreement between experiment and simulation for graphene superlattice devices (see Methods), and the present purely theoretical work can be regarded as a guide for further experimental magneto-transport studies.

\section*{Results}
\paragraph*{No external magnetic field}
We first simulate the four-point longitudinal resistance $R_{xx}$. 
We consider a four terminal device shown in the right inset of \autoref{fig:fig1}(c), where the system length $L=1152$ nm, width $W=385$ nm, and the top lead width $w=245$ nm.
With the four leads labeled in the right inset of \autoref{fig:fig1}(c), we calculate $R_{xx}=R_{14,23}$ (for details see Methods) and show its dependence on the top and back gate voltages in the main panel of \autoref{fig:fig1}(c).
As can be seen from the map, the strength of the backgate mostly influences the superlattice modulation.

\autoref{fig:fig1}(d) presents the linecuts of $R_{xx}$ marked in \autoref{fig:fig1}(c) with the respective colors. Whereas at $V_\mathrm{bg}\approx 0$ only a single Dirac peak is visible, for increasing $|V_\mathrm{bg}|$ second and higher order satellite Dirac peaks start to appear, as the periodic modulation gets stronger. At $V_\mathrm{bg}=\pm 40$ V [linecuts in \autoref{fig:fig1}(d)], a few higher-order Dirac points are resolved. On the other hand, changing the top gate voltage mostly tunes the carrier density in the device. The left inset of \autoref{fig:fig1}(c) shows a close-up of the boxed region with $30\ \mathrm{V} \leq V_\mathrm{bg} \leq 45$ V and $-3.5\ \mathrm{V} \leq V_\mathrm{tg} \leq -1.6$ V, where two sharp lines are visible at both sides of the main Dirac peak, corresponding to the secondary Dirac points, and several fainter lines, corresponding to higher order Dirac points. Similar results based on the same capacitance model function have been reported in \cite{Chen2020}, where two-terminal transport simulations were performed.
The character of the bands can be verified in magnetotransport, as shown in the following subsections.

\paragraph*{Low magnetic field}
%\subsection{Low magnetic field}
 
%\subsubsection{ Semiclassical description of the carrier motion in lattices}

\begin{figure}[t]
\includegraphics[width=\columnwidth]{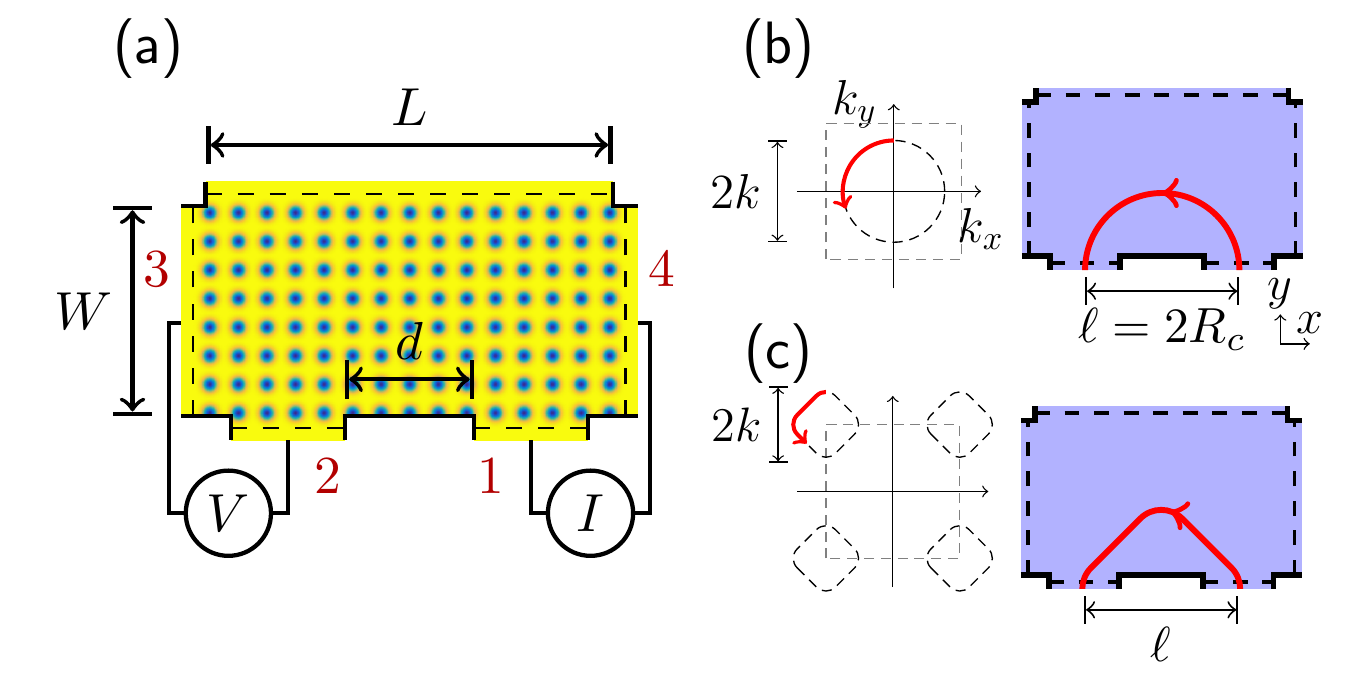} 
\caption{
\textbf{The TMF schematics.}
(a) Sketch of the geometry considered for magnetic focusing. (b)--(c) Fermi contours in the reciprocal space for a circular (b) and noncircular (c) orbit, and the corresponding focused trajectories in real space.
 }
\label{fig:fig2}
\end{figure}

For a general band dispersion $\epsilon(\mathbf{k})$, the semiclassical equations of motion for an electron are given by  \cite{Ashcroft1976}
\begin{align}
\mathbf{\dot r} &= \frac{1}{\hbar} \nabla_\mathbf{k}\epsilon(\mathbf{k}) ,\\ %\frac{\partial}{\partial \mathbf{k}}\epsilon(\mathbf{k}), \\
\hbar \mathbf{\dot k} &= -e(\mathbf{E} + \mathbf{\dot r} \times \mathbf{B}),
\end{align}
where $-e$ is the electron charge, $\mathbf{E}$ is the external electric field, and $\mathbf{B}$ the magnetic field.
In the presence of constant out-of-plane magnetic field $\mathbf{B}=(0,0,B)$ only, one can obtain the relation between the shape of the Fermi contour in the momentum space $\Delta \mathbf{k}(t)$ and the carrier trajectory in real space $\Delta \mathbf{r}(t)$ \cite{Ashcroft1976}
\begin{eqnarray}
\label{eq:semiclassical_dr}
\Delta \mathbf{r}(t) = \frac{\hbar}{eB} \Delta \mathbf{k}(t) \times \hat z,
\end{eqnarray}
meaning that the cyclotron orbit is obtained by rotating the orbit in the momentum space by $90^\circ$ clockwise, as illustrated in \autoref{fig:fig2}(b)--\autoref{fig:fig2}(c). Carriers encircle closed orbits of electron type or hole type in the counterclockwise or clockwise direction, respectively, as determined via the group velocity, $\mathbf{v} = \nabla_\mathbf{k}\epsilon(\mathbf{k}) / \hbar$, and the equation $\hbar \mathbf{\dot k} = -eB \mathbf{v} \times \hat z$. In pristine graphene %or 2D electron gas in semiconductor heterostructures
at low energy, cyclotron orbits exhibit a circular shape [\autoref{fig:fig2}(b)], but the bands formed in systems with SL modulation can be highly distorted from the original, conical shape, and thus, noncircular Fermi contours can be observed [e.g. \autoref{fig:fig2}(c)].

%\subsubsection{Transverse magnetic focusing}
In a typical device designed for transverse magnetic focusing measurement, an emitter and collector [in \autoref{fig:fig2}(a) marked as 1 and 2, respectively] are located on the same edge at a center to center distance $\ell = d+w$, and other contacts act as absorbers.
The current injected from an emitter flows along cyclotron orbits with a radius $R_\mathrm{c}$ depending on the magnetic field strength, and at the boundary propagates along skipping orbits. The current can end up in the collector, when the diameter or its multiples match $\ell$ [\autoref{fig:fig2}(b)], or otherwise in absorber contacts. In the nonlocal resistance measurement [\autoref{fig:fig2}(a)], this results in maximum or minimum, respectively, of the resistance $R_\mathrm{f} = R_{14,23}$ (see Methods). %see Eq.\ (\ref{eq:Rklmn}). 
For electron-like (hole-like) orbits, the resistance maximum condition
can be obtained for positive (negative) $B$. 

 In pristine graphene, a typical TMF spectrum as a function of magnetic field and voltage contains two fans of focusing peaks, one for the electron band and the other for the hole band \cite{Taychatanapat2013}. In a superlattice, the emerging replicas of the Dirac cone cause a substantial modification of the TMF signal. In the following discussion, we choose $V_\mathrm{bg}=40$ V, such that a few higher-order Dirac peaks are present next to the main Dirac point as seen in \autoref{fig:fig1}(c).

\begin{figure*}[t]
\includegraphics[width=\textwidth]{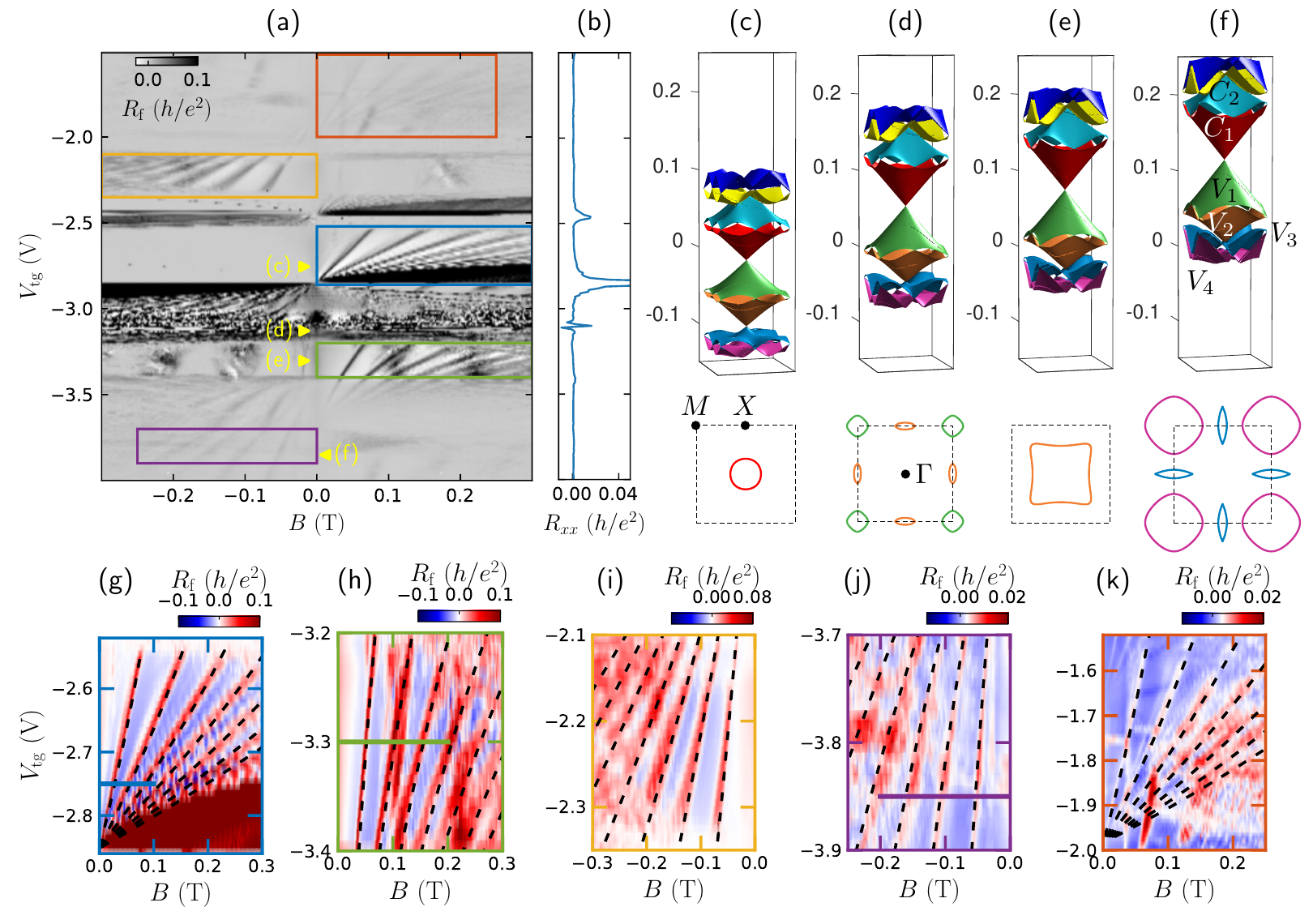} 
\caption{
\textbf{Magnetic focusing in the superlattice.}
(a) Nonlocal resistance $R_\mathrm{f}$ as a function of magnetic field and top gate voltage at $V_\mathrm{bg} = 40$ V.
(b) $R_{xx}$ calculated for the same system at $B=0$.
(c)--(f) Top row: the miniband structures calculated at $V_\mathrm{tg}$ marked by the arrows in (a). In (f) $C_1$, $C_2$, and $V_1$--$V_4$ label selected minibands. Bottom row: the corresponding Fermi contours at $E=0$. 
(g)--(k) The zoomed map in (a) at the rectangles marked with the corresponding colors.
The dashed lines mark the focusing condition $\ell/2n=\hbar k/eB$, $n=1,2,\dots$. 
}
\label{fig:fig3}
\end{figure*}

For transverse magnetic focusing, we choose the system geometry shown in \autoref{fig:fig2}(a), with the distance between the bottom leads $d=1200$ nm, their widths $w=100$ nm, the side leads width $W=1636$ nm, and the top lead width $L=1792$ nm.

\autoref{fig:fig3}(a) shows the $R_\mathrm{f}$ map as a function of $B$ and $V_\mathrm{tg}$. One can see several series of fans, with the focusing peaks appearing at $B>0$ or $B<0$. 
The map is put together with the $R_{xx}$ calculated at $B=0$, shown in \autoref{fig:fig3}(b), where the main and secondary Dirac peaks are seen. The sign change of the focusing peaks in \autoref{fig:fig3}(a) occurs either at the Dirac points, or van Hove singularities, and for the former, it coincides with the $R_{xx}$ peaks. 
This confirms the sign change of the carriers when tuning $V_\mathrm{tg}$, occurring as an effect of the band reconstruction due to the superlattice potential.
To understand the result in detail, in \autoref{fig:fig3}(c)--\autoref{fig:fig3}(f) we plot the miniband structures calculated at $B=0$, as described in Ref.\ \cite{Chen2020}, and the Fermi contours at $E=0$, at selected values of $V_\mathrm{tg}$ marked by arrows in \autoref{fig:fig3}(a). Additionally, in \autoref{fig:fig3}(g)--\autoref{fig:fig3}(k) we show zoomed regions of the $R_\mathrm{f}$ map marked with the colored rectangles in \autoref{fig:fig3}(a).

In \autoref{fig:fig3}(c), at $V_\mathrm{tg}=-2.75$ V, the Fermi level is located at the $C_1$ subband, and the Fermi contour has a rounded shape [\autoref{fig:fig2}(b) and \autoref{fig:fig3}(c)]. In the semiclassical description,
  electrons injected from lead 1 [\autoref{fig:fig2}(a)] with an initial velocity $\mathbf{v}=(0,v)$ and $\mathbf{k}=(0, k)$, in a moderate magnetic field travel along a rounded trajectory, and after half a period, $t=T/2$ encircle half of the closed orbit.
From Eq.\ (\ref{eq:semiclassical_dr}), this corresponds to traveling a distance equal to the diameter, $2R_\mathrm{c} = 2\hbar k/eB$ along the $x$ direction [\autoref{fig:fig2}(b)]. For $\ell=2R_\mathrm{c}$, it leads to the first maximum of $R_\mathrm{f}$. For smaller $R_\mathrm{c}$, the beam is reflected at the edge, and can flow to the collector when $\ell=4R_\mathrm{c}, 6R_\mathrm{c}, \dots$, giving rise to higher order $R_\mathrm{f}$ peaks. In general, we can evaluate the field at which the $j$th maximum occurs as $B_j=2\hbar j k/e\ell$, $j=1,2,\dots$. 
We find $k(V_\mathrm{tg})$ numerically and plot $B_j$ with dashed lines in \autoref{fig:fig3}(g). The $C_1$ subband cone is within $-2.9\ \mathrm{V}\lesssim V_\mathrm{tg}\lesssim -2.5$ V. We find a very good agreement with the $R_\mathrm{f}$ signal for up to $j\approx 7$. Higher $j$ are not resolved as the system enters the quantum Hall regime, and semiclassical description of the skipping orbits at the edge no longer applies. At $-3.1\ \mathrm{V}\lesssim V_\mathrm{tg}\lesssim -2.9$ V, for the $V_1$ subband, $R_\mathrm{f}$ is noisy due to scattering of low-energy carriers by the periodic potential. 

When the Fermi level is tuned to the van Hove singularity (at $V_\mathrm{tg}\approx -2.5$ for the electron subband, and $V_\mathrm{tg}\approx -3.1$ for the hole subband), the focusing signal vanishes, and smaller fans reappear. 
Based on the miniband structures [\autoref{fig:fig3}(d)], we interpret them as due to focusing of the secondary Dirac cones fermions. For example, at $V_\mathrm{tg}=-3.12$ V [\autoref{fig:fig3}(d)] at $E=0$ there are tiny Dirac cones around the $M$ and $X$ points of the Brillouin zone. 

For $V_\mathrm{tg}\lesssim -3.2$ V and $V_\mathrm{tg}\gtrsim -2.4$ V, the miniband structures around the Fermi level get more complex, with many overlapping subbands. Nevertheless, we find ranges of $V_\mathrm{tg}$ where a single isolated higher-order Dirac cone is present, giving clear focusing signal [see \autoref{fig:fig3}(e) for the $V_2$ subband, the corresponding $R_\mathrm{f}$ zoom in \autoref{fig:fig3}(h), and the zoom in \autoref{fig:fig3}(i) for the $C_2$ subband]. When there are more overlapping subbands, the signal gets very faint.  Nevertheless, one can spot fans that fit well to the hole-like orbit within the $V_4$ subband around the $M$ point, see \autoref{fig:fig3}(f) at $V_\mathrm{tg}=-3.85$ V, and zoomed $R_\mathrm{f}$ in \autoref{fig:fig3}(j). A similar feature is resolved in \autoref{fig:fig3}(k) for the electron-like orbits.

\begin{figure}[t]
\includegraphics[width=\columnwidth]{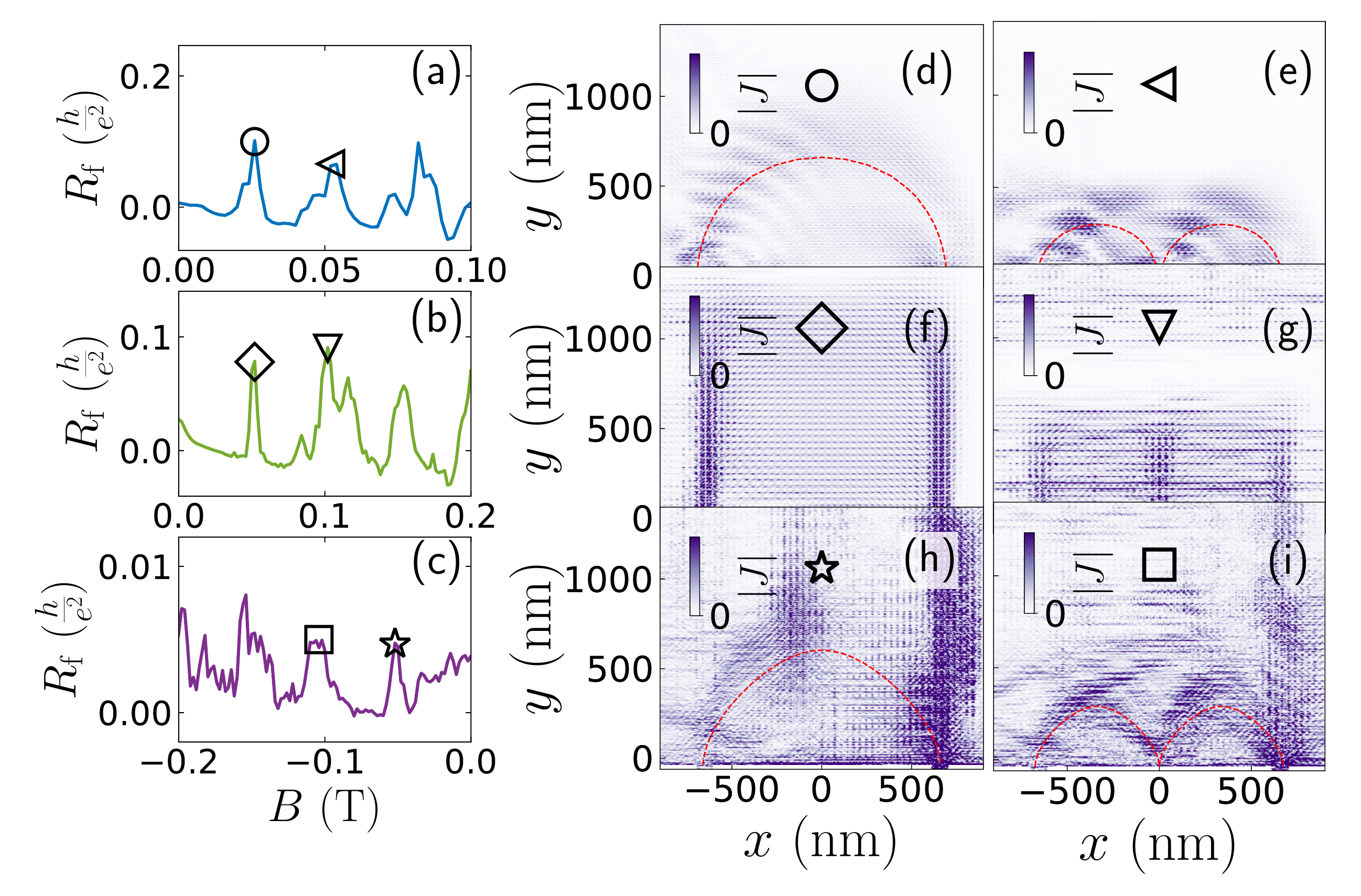} 
\caption{
\textbf{Non-circular motion of Dirac fermions.}
Line cuts at $V_\mathrm{tg}$ and $B$ ranges as marked in \autoref{fig:fig3}(g), \autoref{fig:fig3}(h), and \autoref{fig:fig3}(j) by the respective colors at (a) $V_\mathrm{tg}=-2.75$ V, (b) $V_\mathrm{tg}=-3.3$ V, and (c) $V_\mathrm{tg}=-3.85$ V. (d) -- (i) are the current density maps in arbitrary units at $B$ marked in (a) -- (c) by the respective symbols. The red dashed lines show the expected semi-classical trajectory.
} 
\label{fig:fig4}
\end{figure}

To further illustrate the relation of the real-space orbits to the subbands, we calculate the current density maps at selected focusing peaks. 
\autoref{fig:fig4}(a)--\autoref{fig:fig4}(c) show the line cuts of $R_\mathrm{f}$ at $V_\mathrm{tg}$ and $B$ marked by the respective colors in \autoref{fig:fig3}(g), \autoref{fig:fig3}(h) and \autoref{fig:fig3}(j). In \autoref{fig:fig4}(a) for the $C_1$ subband, the first two focusing peaks are marked by {\large $\circ$} and $\triangleleft$. The respective current density maps are presented in \autoref{fig:fig4}(d) and \autoref{fig:fig4}(e), revealing typical current densities found for TMF calculations in pristine graphene \cite{Stegmann2015, Petrovic2017}. 
In \autoref{fig:fig4}(b) the line cut for the $V_2$ subband is shown. For the focusing peaks marked with {\large $\diamond$} and $\triangledown$, the current density maps are shown in \autoref{fig:fig4}(f) and \autoref{fig:fig4}(g). The trajectories acquire a shape close to a rectangle, consistent with the Fermi contour [\autoref{fig:fig3}(e)].
For the line cut in the hole-like $V_4$ subband (purple) [\autoref{fig:fig4}(c)], the current densities of the first two peaks are shown in \autoref{fig:fig4}(h) and \autoref{fig:fig4}(i)]. The orbits acquire a rhombus shape, matching well the Fermi contour in the corner of the Brillouin zone [\autoref{fig:fig2}(c) and \autoref{fig:fig3}(f)]. 
The red dashed lines show semiclassical trajectories calculated using the Fermi contours obtained from our band structures and \autoref{eq:semiclassical_dr} with the contour starting at $\mathbf{k}$ for which $v_x \propto \partial E/\partial k_x=0$. These semiclassical orbits show similarity with the current density, but the current density is obtained from quantum calculation so they are expected to be similar but not strictly identical, in particular, counter-intuitive patterns may occur at certain resonant conditions (e.g. a vertical blob on top of the rhombus-like pattern in \autoref{fig:fig4}(h)). Note that in \autoref{fig:fig4}(f)--\autoref{fig:fig4}(g) the current density is non-zero in the area between the vertical segments, as it contains contributions from multiple initial $k_x$ for which $v_x$ is low. The noisy background visible in \autoref{fig:fig4}(a)--\autoref{fig:fig4}(c) originates from scattering and resonant states due to SL which are irregular and complex due to the wave-like nature of the carriers. 

%\subsection{High magnetic field}

\begin{figure*}[t]
\includegraphics[width=\textwidth]{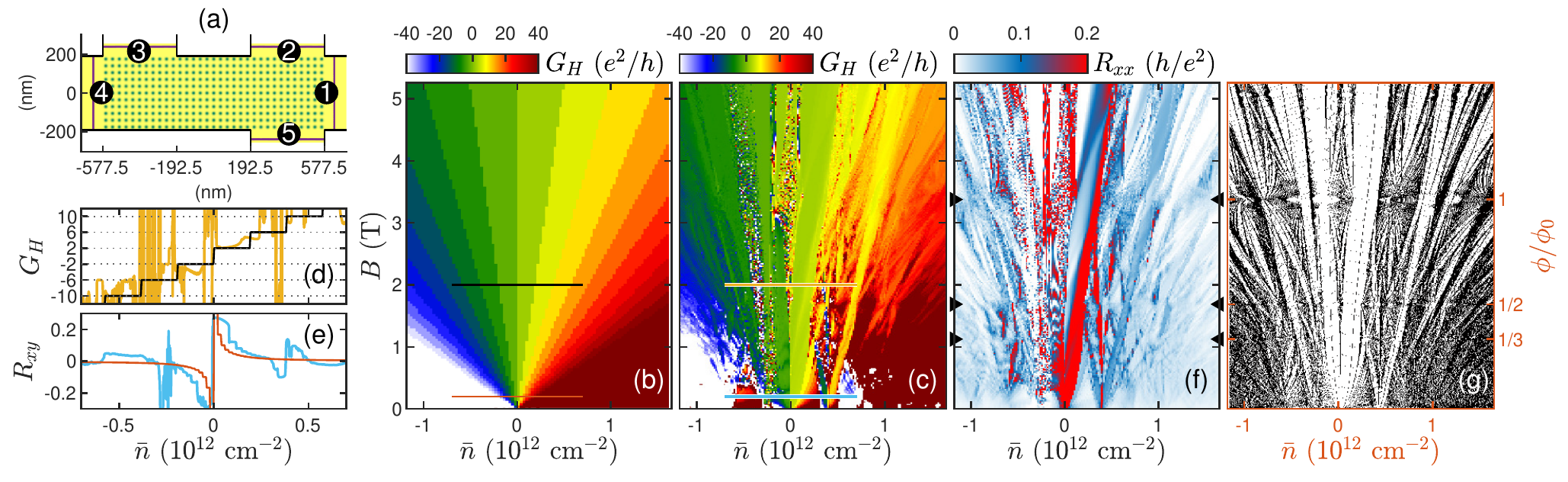} 
\caption{
\textbf{High-field transport.}
(a) The 5-terminal system used for the calculation. (b) and (c) are $G_\mathrm{H}=R_{xy}^{-1}$ as a function of average carrier density $\bar{n}$ and magnetic field $B$ with $V_\mathrm{bg}$ fixed at $0$ and $40\unit{V}$, respectively, where horizontal lines at $B=2\unit{T}$ and $B=0.2\unit{T}$ mark the line cuts shown in (d) for $G_\mathrm{H}$ and (e) for $R_{xy}$. (f) Longitudinal resistance $R_{xx}$ as a function of $\bar{n}$ and $B$. (g) Calculated Hofstadter spectrum corresponding to (c) and (f).} 
\label{fig:fig5}
\end{figure*}

%We limit the detailed discussion to $V_\mathrm{tg}$ below the main Dirac point. 
Although the focusing spectrum is not symmetric with respect to the main Dirac point, it is qualitatively similar in minibands above and below the main Dirac point, except for the noisy signal for low-energy valence subbands.
Let us note that the modulation induced by electrostatic gates is a complex function of $V_\mathrm{bg}$ and $V_\mathrm{tg}$, and the band structure is significantly modified merely by changing $V_\mathrm{tg}$ [\autoref{fig:fig3}(c)--\autoref{fig:fig3}(f)]. This is in contrast to the SLs induced via low-angle twisted hBN substrate or twisted graphene layers, where the top gate only sweeps the carrier density without affecting the periodic modulation, and the band structure remains unchanged. 
This leads to the shape of the fans in $R_\mathrm{f}$ %around the main Dirac point being 
closer to straight lines, unlike Refs.\ \cite{Lee2016, Berdyugin2020} that observed ones close to parabolic.

\paragraph*{High magnetic field}
Now, we turn our attention to the high-field regime, where we expect the Hofstadter spectrum to emerge when the magnetic flux per SL unit cell of area $A$, $\phi=AB$, is of the order of the flux quantum $\phi_0=h/e$. To simulate longitudinal resistance $R_{xx}$ and Hall resistance $R_{xy}$ at the same time, while keeping calculations for the four-point resistances minimized, we consider a 5-terminal Hall bar sketched in \autoref{fig:fig5}(a) with the actually considered geometric dimensions shown. We compute all the $5\times 4=20$ transmission functions between all pairs among the five leads, and process the data to obtain longitudinal resistance $R_{xx}=R_{14,23}$ and Hall resistance $R_{xy}=R_{14,52}$ following the B\"uttiker formalism \cite{Buttiker1986, Datta1995}, according to the lead labels shown in \autoref{fig:fig5}(a). For all the following discussions, we convert our $V_\mathrm{tg}$ axis to the numerically obtained average carrier density $\bar{n}$ over the entire lattice, in order for a more transparent presentation of our high-field transport simulations.

\autoref{fig:fig5}(b) and \autoref{fig:fig5}(c) show the Hall conductance $G_\mathrm{H}=R_{xy}^{-1}$ as a function of $\bar{n}$ and magnetic field $B$, with the back gate voltage fixed at $V_\mathrm{bg}=0$ and $V_\mathrm{bg}=40\unit{V}$, respectively. To better highlight the Landau fans arising from the quantum Hall effect of graphene, we limit the color range to $-40\leq G_\mathrm{H}/G_0\leq +40$ in \autoref{fig:fig5}(b), where $G_0=e^2/h$ is the conductance quantum. The same limit of the color range is applied to \autoref{fig:fig5}(c) in order for a direct comparison. Line cuts at $B=2\unit{T}$ marked by the black line on \autoref{fig:fig5}(b) and orange line on \autoref{fig:fig5}(c) are shown and compared in \autoref{fig:fig5}(d). The lowest few quantum Hall conductance plateaus $G_\mathrm{H}=\pm 2G_0,\pm 6G_0,\pm 10 G_0, \cdots$ can be clearly seen in the $V_\mathrm{bg}=0$ case free of superlattice potential (black line), while the combined strong magnetic field ($B=2\unit{T}$) and strong superlattice modulation ($V_\mathrm{bg}=40\unit{V}$) result in a more complex transport feature, in accordance with the predictions \cite{Thouless1982, Streda1982} for a periodic 2D modulation, which can be better understood by checking $R_{xx}$, instead of $R_{xy}$ (or $G_\mathrm{H}$), to be discussed soon below.

Without taking the inverse of $R_{xy}$, \autoref{fig:fig5}(e) shows the Hall resistance at $B=0.2\unit{T}$ with $V_\mathrm{bg}=0$ [red, corresponding to the $\bar{n}$ range marked on \autoref{fig:fig5}(b)] and $V_\mathrm{bg}=40\unit{V}$ [cyan, corresponding to the $\bar{n}$ range marked on \autoref{fig:fig5}(c)]. The former (without superlattice) shows a single sign change at $\bar{n}=0$, typical for graphene, while the latter (with superlattice) shows multiple sign changes at positive and negative $\bar{n}$, in addition to the main Dirac point at $\bar{n}=0$, consistent with our previous low-field magnetotransport simulations discussed above. %in [{\color{red} cite or refer}].

Considering the same range of $\bar{n}$ and $B$ as \autoref{fig:fig5}(b) and \autoref{fig:fig5}(c), \autoref{fig:fig5}(f) shows the longitudinal resistance $R_{xx}$ with the back gate voltage fixed at $V_\mathrm{bg}=40\unit{V}$. Since the period of our gate-controlled superlattice is $\lambda=35$ nm, and hence the square SL unit cell area $A=\lambda^2=1225\unit{nm^2}$, the condition $\phi/\phi_0=AB/(h/e)=1$ is reached at $B\approx 3.376\unit{T}\equiv B_1$, which is in sharp contrast with the graphene/hBN moir\'e superlattice that requires $B\approx 24\unit{T}$ in order to reach $\phi/\phi_0=1$, because of its periodicity limited to $\lambda\lesssim 14\unit{nm}$ and hence the area $A=\sqrt{3}\lambda^2/2\lesssim 170\unit{nm^2}$. As is visible on \autoref{fig:fig5}(f), the seemingly complicated $R_{xx}$ map does exhibit certain features at $B_1$ and $B_1/2$, and vaguely at $B_1/3$ (marked by black triangles), corresponding to $\phi/\phi_0=1,1/2,1/3$, respectively, consistent with our calculation of the magnetic energy subbands shown in \autoref{fig:fig5}(g), where the vertical axis of $\phi/\phi_0$ corresponds to exactly the same $B$ range as \autoref{fig:fig5}(f), as well as the average density range. Good consistency between the $R_{xx}$ map of \autoref{fig:fig5}(f) and the Hofstadter butterfly shown in \autoref{fig:fig5}(g) can be seen. For methods adopted to calculate \autoref{fig:fig5}(g), see Methods.

\section*{Discussion}
In summary, we theoretically investigated transport in gated superlattices based on monolayer graphene. Our zero and low magnetic field transport calculations remain in a good agreement with the continuum model band structure calculated in presence of periodic modulation. 
We explored the potential of TMF for probing the intricate band structure of graphene with periodic modulation. 
It offers possibilities to study a plethora of phenomena in superlattices,  
and opens the door for studies of strongly correlated systems in twisted bilayer graphene \cite{deVries2021} or in bilayer graphene superlattices \cite{Krix2022}. 
By exploring the reconstructed band structure via magnetotransport calculations it is possible to engineer devices relying on directed electron flow due to the distortion of Fermi contour, as well as for other applications based on mini band electron optics. 
We also obtained high-magnetic-field response consistent with the Hofstadter spectrum calculated for a gated superlattice as a function of the gate voltage.
Our modeling can be generalized to other superlattice geometries, and is promising for the investigations of future band structure engineered devices working in a broad range of magnetic fields.

\section*{Methods}

\paragraph*{Gated superlattice model}
To model a realistic geometry, we consider SL induced by a patterned dielectric substrate \cite{Forsythe2018, Li2021} with a uniform global backgate underneath the dielectric layer, and the graphene sheet sandwiched between two hBN layers lying on top [\autoref{fig:fig1}(a) of the main text]. The hBN/graphene/hBN sandwich is covered by a global top gate. The voltage applied to the back gate $V_{\mathrm{bg}}$ controls the strength of the periodic modulation, while the top gate voltage $V_{\mathrm{tg}}$ is used to tune the carrier density across the SL. We follow the design of Ref.\ \cite{Forsythe2018}, with a square lattice etched in SiO$_2$ substrate, with a lattice period $\lambda=35$ nm. We use a model function $C_{\mathrm{bg}}(x,y)$ 
%\begin{equation}
\begin{align}
C_{\mathrm{d}}(x,y) &= C_{\mathrm{out}} - (C_{\mathrm{out}} - C_{\mathrm{in}}) \tanh\left[ \exp\left(- \frac{x^2 + y^2}{ d_{\mathrm{smooth}}^2}\right)\right],\\
\label{eq Cbg model function}
C_{\mathrm{bg}}(x,y) &= C_{\mathrm{d}}\left[ \left(x - \frac{\lambda}{2}\right) \% \lambda - \frac{\lambda}{2}, \left(y - \frac{\lambda}{2}\right) \% \lambda - \frac{\lambda}{2}\right],
\end{align}
%\end{equation}
where $ d_{\mathrm{smooth}} = 7.5$ nm is the smoothness of the modulation, $C_{\mathrm{out}}/e=0.77\times 10^{11}\ \mathrm{cm}^{-2}\mathrm{V}^{-1}$ and $C_{\mathrm{in}}/e=0.22\times 10^{11}\ \mathrm{cm}^{-2}\mathrm{V}^{-1}$ , and $a\%b$ means the remainder after dividing $a$ by $b$. 
The top gate capacitance is assumed to be $C_{\mathrm{tg}}/e=1\times 10^{12}\ \mathrm{cm}^{-2}\mathrm{V}^{-1}$. Using the parallel capacitor model, this roughly corresponds to the top hBN thickness $d_\mathrm{t}\approx 16.6$ nm, from $C_\mathrm{tg}/e = \varepsilon_0\varepsilon_\mathrm{hBN}/e d_\mathrm{t}$, where $\varepsilon_0$ is the vacuum permittivity, $\varepsilon_\mathrm{hBN}=3$ is the dielectric constant of hBN, and $-e$ is the electron charge. Figure 1(b) of the main text shows the profile of the above model function (\ref{eq Cbg model function}).

For the dual-gated graphene sample free from intrinsic doping, the carrier density is given by 
\begin{equation}
\label{eq:SLn}
n(x,y) = \frac{C_\mathrm{bg}(x,y)}{e}V_\mathrm{bg} + \frac{C_\mathrm{tg}}{e}V_\mathrm{tg}.
\end{equation}
Assuming that the carrier energy in graphene is given by $E=\pm \hbar v_\mathrm{F} k$, where $\hbar$ is the reduced Planck constant, $v_\mathrm{F}\approx 10^6\unit{m/ s}$ is the Fermi velocity of graphene, %$k=\sqrt{\pi n}$ is the wavenumber, 
and using $\hbar v_\mathrm{F} \approx 3\sqrt{3}/8\ \unit{eV nm}$, the onsite potential energy can be obtained from
\begin{equation}
\label{eq:SLE}
U=-\sgn(n) \hbar v_\mathrm{F} \sqrt{\uppi |n|},
\end{equation}
%and is used in the transport calculations.
in order to set the global Fermi energy at $E = 0$ where transport occurs.

\paragraph*{Transport calculation}
For transport calculation, we use the tight-binding Hamiltonian 
\begin{equation}
\label{eq:Htb}
H = -\sum\limits_{\left\langle {i,j} \right\rangle } t_{ij} c_i^\dagger  c_j + \sum\limits_j {U({\mathbf{r}}_j)} c_j^\dagger c_j,
\end{equation}
where $c_i$ ($c_i^\dag$) is an annihilation (creation) operator of an electron on site $i$ located at $\mathbf{r}_i=(x_i,y_i)$. The first sum contains the nearest-neighbor hoppings with the hopping parameter $t_{ij}$, and the second sum describes the onsite potential energy profile. 
In the presence of an external magnetic field $\mathbf{B} = (0,0,B)$, the hopping integral is modified by $t_{ij} \rightarrow t_{ij}\exp(i\phi)$, where the Peierls phase $\phi=(-e/\hbar) \int_{\textbf{r}_i}^{\textbf{r}_j} \textbf{A}\cdot d\textbf{r}$, with $\textbf{A}$ being the vector potential that satisfies $\nabla\times \mathbf{A} = \mathbf{B}$, and the integral going from the site at $\textbf{r}_i$ to the site at $\textbf{r}_j$.
For a feasible simulation of realistic devices, we use the scalable tight-binding model \cite{Liu2015}, where the hopping parameter becomes $t_{ij} = t_0/s_\mathrm{f}$, and the lattice spacing $a = a_0s_\mathrm{f}$, $s_\mathrm{f}$ is the scaling factor, and we use $t_0=-3$ eV and $a_0 = 1/4\sqrt{3}$ nm. 
Transport calculations based on Hamiltonian (\ref{eq:Htb}) are done within the wave-function matching for the TMF, and real-space Green's function method in other cases, at the global Fermi energy $E=0$ and zero temperature.
%The transmission probability is evaluated 
The conductance from lead $i$ to lead $j$ is obtained from the Landauer formula $G_{ji}= 2(e^2/h) T_{ji}$, where the transmission probability $T_{ji}$ is evaluated as a sum over the propagating modes $T_{ji} = \sum\limits_{q} T_{ji}^{q}$, and
\begin{equation}
T_{ji}^{q} = \sum\limits_{p} |t_{ij}^{pq}|.
\end{equation}
Here, $t_{ij}^{pq}$ is the probability amplitude for the transfer from the incoming mode $p$ in lead $i$ to the outgoing mode $q$ in lead $j$. 

In the multiterminal devices, we solve the transport problem for each lead as an input, and build the conductance matrix $\mathbfcal{G}$ \cite{Buttiker1986, Datta1995} which relates the current $I_i$ fed to the system in lead $i$ to the voltage at $j$-th lead $V_j$ through $I_i = \sum_{j=1}^N \mathcal{G}_{ij} V_j$. For an $N$-terminal system, the matrix elements are 
\begin{align}
\mathcal{G}_{ij} &= -G_{ij},\ \quad i\ne j, \\
\mathcal{G}_{ii} &= \sum\limits_{j=1,j\ne i}^N G_{ij}.
\end{align} 
We set the voltage at $l$-th lead equal to zero, and eliminate the $l$-th row and column of the matrix. The reduced $(N-1)\times(N-1)$ matrix $\bar{\mathbfcal{G}}$ can be inverted to get $\boldsymbol{\mathbfcal{R}} = \bar{\mathbfcal{G}}^{-1}$, where the $\mathbfcal{R}$ matrix satisfies 
\begin{equation}
\label{eq:Rmatrix}
%V_{i} = \sum_{j=1}^{N-1} \mathcal{R}_{ij} I_j.
V_{i} = \sum_{j=1,j\ne l}^{N} \mathcal{R}_{ij} I_j.
\end{equation}
With the elements of matrix $\boldsymbol{\mathbfcal{R}}$, one can evaluate the resistance 
\begin{equation}
%R_{ij,kl} = \left. \frac{V_{k} - V_{l}}{I_i} \right|_{I_j=-I_i}
% = \frac{\sum_{n=1}^{N-1} \mathcal{R}_{kn} I_n - \sum_{n=1}^{N-1} \mathcal{R}_{ln} I_n}{I_i}
% = \frac{ \mathcal{R}_{ki} I_i + \mathcal{R}_{kj} I_j - \mathcal{R}_{li} I_i - \mathcal{R}_{lj} I_j}{I_i}
% = \mathcal{R}_{ki} - \mathcal{R}_{kj} - \mathcal{R}_{li} + \mathcal{R}_{lj} 
%R_{iN,kl} = \frac{V_{k} - V_{l}}{I_i}  = \mathcal{R}_{ki} - \mathcal{R}_{li} 
\label{eq:Rklmn} R_{kl,mn} = \frac{V_{m} - V_{n}}{I_k} = \mathcal{R}_{mk} - \mathcal{R}_{nk} 
\end{equation}
with the current flowing from lead $k$ to lead $l$, zero current in other terminals, and voltage drop measured between leads $m$ and $n$.

\begin{figure}
%\centering
\includegraphics[width=\textwidth]{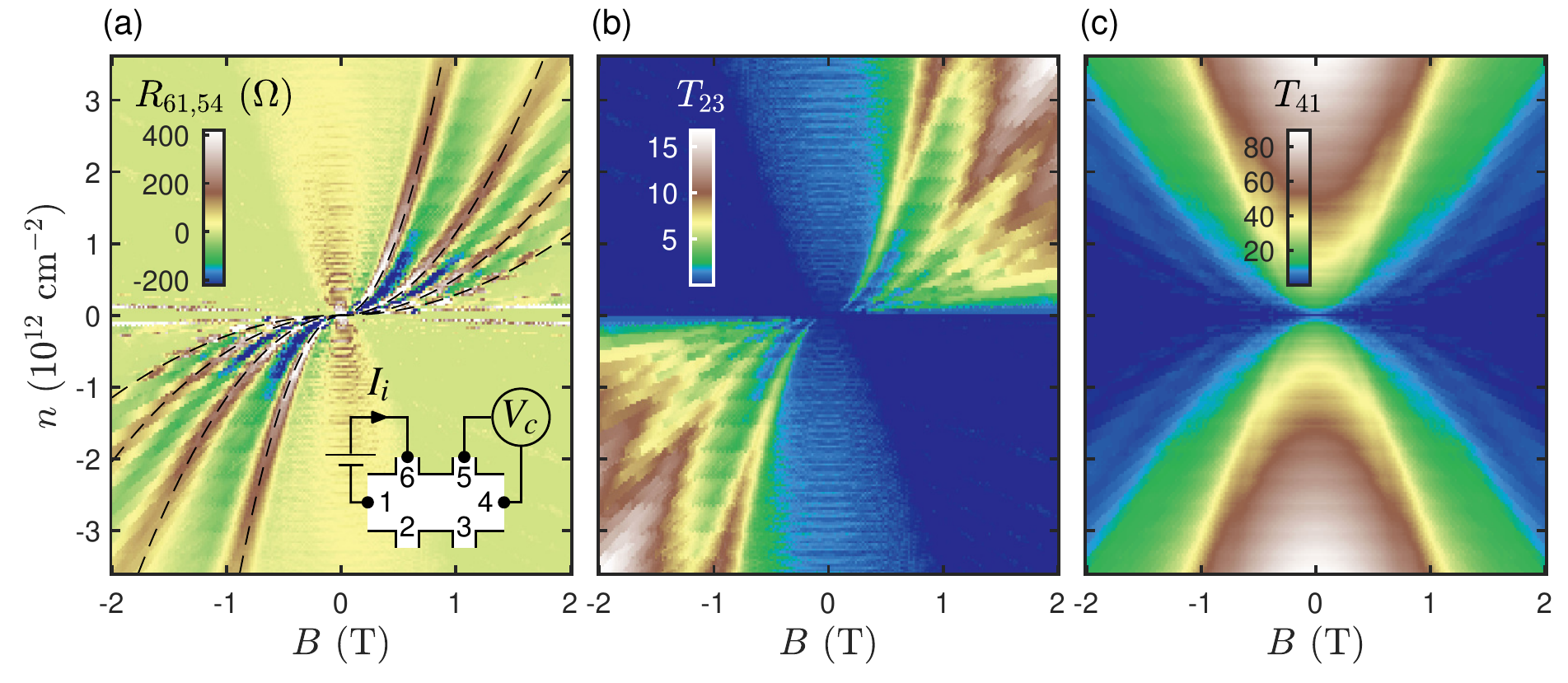}
\caption{
\textbf{Revisiting the TMF experiment on graphene.}
(a) Computed four-point resistance $R_{61,54}$ as a function of magnetic field $B$ and carrier density $n$ revisiting the TMF experiment on graphene \cite{Taychatanapat2013}, and two exemplary transmission functions required for $R_{61,54}$: (b) $T_{23}$ and (c) $T_{41}$. The inset in (a) shows the orientation of the considered Hall bar and the configuration of the leads for ground, injector, and voltage probes.}
\label{fig:TMF}
\end{figure}

\paragraph*{Transverse magnetic focusing}
\label{sec:tmf}
As a numerical example of applying the above outlined Landauer-B\"uttiker formalism for computing the four-point resistance Eq.\ \eqref{eq:Rklmn}, we revisit the first TMF experiment on graphene \cite{Taychatanapat2013}, considering the same probe spacing ($\ell=500$~nm) and width ($100$~nm) but slightly different geometry of the scattering region (total length $700$~nm and width $500$~nm) for a 6-terminal Hall bar, made of a graphene lattice scaled by $s_\mathrm{f}=12$. Choosing the same configuration of the leads for injector, ground, and voltage probes as the revisited experiment, the computed $R_{61,54}$ as a function of the external magnetic field $B$ perpendicular to the plane of graphene and the carrier density $n$ is reported in \autoref{fig:TMF}(a), showing a map rather consistent with the experiment. Due to the isotropic low-energy dispersion of graphene, the resulting cyclotron trajector is a simple circle of radius $R_\mathrm{c}=\hbar k/eB$, which is simplified from Eq.\ \eqref{eq:semiclassical_dr}. By requiring the probe spacing to be equal to an integer times the cyclotron diameter, $\ell=j\cdot 2R_\mathrm{c}$, $j=1,2,\cdots$, together with $k=\sqrt{\uppi|n|}$ for graphene, one can solve for carrier density corresponding to the $j$th peak of the TMF on the $B$-$n$ map of \autoref{fig:TMF}(a):
\begin{equation}
n_j(B) = \frac{1}{\uppi}\left(\frac{eB\ell}{2\hbar j}\right)^2\ .
\label{eq:n_j for TMF}
\end{equation}
The dashed lines on \autoref{fig:TMF}(a) show $n_1,n_2,n_3,n_4$, matching very well with the patterns of the simulated $R_{61,54}$, which requires totally $6\times 5=30$ transmission functions for such a 6-terminal device, as explained above. Figures \ref{fig:TMF}(b) and (c) show two exemplary maps of transmission functions, which can look generally very different from the resulting four-point resistance.

\paragraph*{Choosing the gauge}
\label{sec:gauge}
In the presence of an external magnetic field and semi-infinite leads, the vector potential must satisfy the translational invariance of the leads. For the magnetic field along the $\hat z$ axis, the most common choice is the Landau gauge, $\mathbf A = (-yB, 0, 0)$ or $\mathbf A = (0, xB, 0)$ for the lead which is translationally invariant along the $x$ or $y$ direction, respectively. For other lead orientation, in general, the proper gauge is different. 
%For a uniform magnetic field in the entire device, the vector potential in the entire device needs to conform with $\nabla\times \mathbf{A} = \mathbf{B}$. 
Therefore, in a multi-terminal device, the required vector potential is not uniform in the entire space. This is not a problem since adding an arbitrary curl-free component to the vector potential does not change the magnetic field. Here, we use the approach introduced in \cite{Baranger1989}.%, where the appropriate gauge is chosen in each lead, and then apply it in the region where the lead is translationally invariant. 

Assuming that the proper gauge in the 1st lead is $\mathbf{A}_1(\mathbf{r})$, for another lead that is at an angle $\theta_n$ with respect to lead 1, the gauge can be chosen as
\begin{equation}
\label{eq:gauge_transformation}
\mathbf{A}_n(\mathbf{r}) = \mathbf{A}_1(\mathbf{r}) + \boldsymbol \nabla f_n(\mathbf{r}),
\end{equation}
with 
\begin{equation}
f_n(\mathbf{r}) = B x y \sin^2(\theta_n) + \frac{1}{4} B (x^2 - y^2) \sin(2\theta_n).
\end{equation}
The addition of a gradient of a scalar function does not influence the requirement $\nabla\times \mathbf{A} = \mathbf{B}$. As an illustration of the transformation, consider $\mathbf{A}_1(\mathbf{r}) = (-yB, 0, 0)$, and $\theta_n=90^\circ$. Then, $f_n(\mathbf{r}) = B x y$,  $\nabla f_n(\mathbf{r}) = (B y, Bx, 0)$, and $\mathbf{A}_n(\mathbf{r})= (-yB,0,0)+(yB,xB,0)  = (0, xB, 0)$.

Applying the transformation (\ref{eq:gauge_transformation}) so that it only affects lead $n$ is possible by defining a smooth step function $\zeta_n(\mathbf{r})$ which is only nonzero in the translationally invariant part of lead $n$
\begin{equation}
\zeta_n(\mathbf{r}) = 
\begin{cases}
1,&	\mathbf{r} \text{ in lead } n, \\
0,&	\mathbf{r} \text{ in lead } m\ne n, \\
\text{smooth interpolation} &\mathbf{r} \text{ elsewhere}.
\end{cases}
\end{equation}
%Note that for this to be possible we may need a buffer layer/area in the lead.
Then, in (\ref{eq:gauge_transformation}) we substitute $f_n(\mathbf{r})\rightarrow \zeta_n(\mathbf{r})f_n(\mathbf{r})$ for lead $n$. In general, for the entire system we define
\begin{equation}
f(\mathbf{r}) = \sum\limits_{n=2}^N \zeta_n(\mathbf{r}) f_n(\mathbf{r}),
\end{equation}
this completes our gauge transformation. Adopting the vector potential
\begin{equation}
\label{eq:gauge_A}
\mathbf{A}(\mathbf{r}) = \mathbf{A}_1(\mathbf{r}) + \boldsymbol \nabla f(\mathbf{r}),
\end{equation}
we have $\mathbf{B} = \nabla \times \mathbf{A}$ everywhere in the system, and the translation invariance in each lead is guaranteed. Importantly, curl of (\ref{eq:gauge_A}) gives exactly the desired $B$, regardless of the smoothness of the $\zeta_n$ function.

\begin{figure}[t]
\begin{center}
\includegraphics[width=8.6cm]{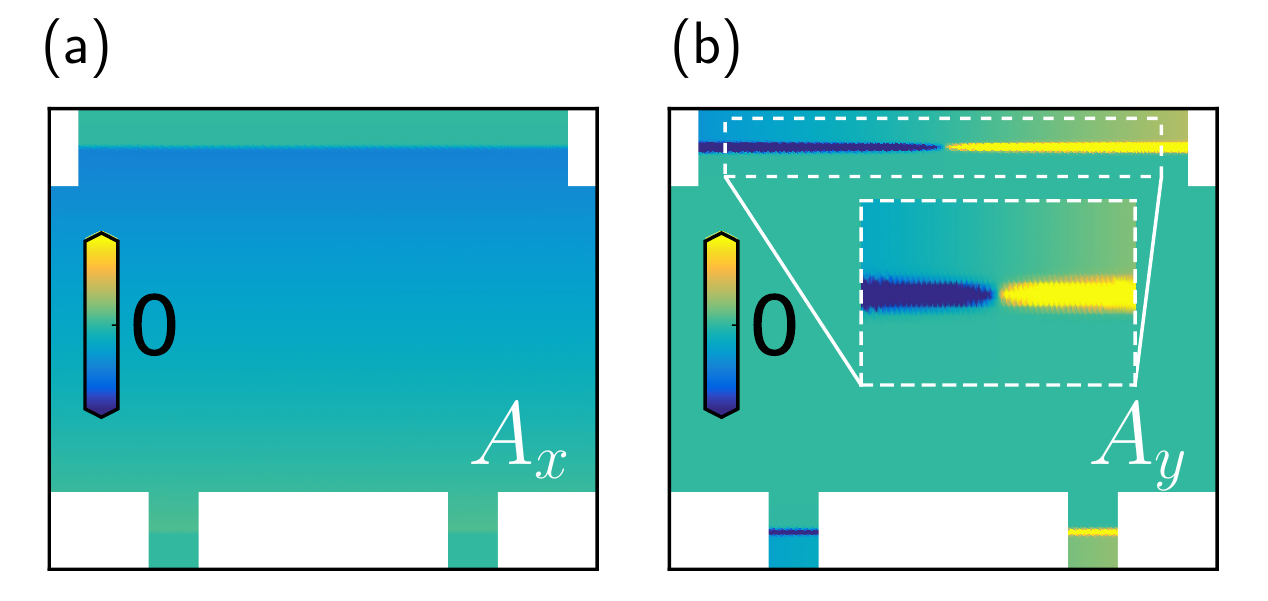} 
\end{center}
\caption{
\textbf{Spatial profiles of the vector potential.}
(a) $A_x$ and (b) $A_y$ vector potential components used for the 5-terminal TMF simulation.
 }
\label{fig:fig6}
\end{figure}

As an example, for the system used for the TMF modeling [\autoref{fig:fig2}(a)], in the vertical leads we choose the same gauge $\textbf{A} = (0, xB)$ with $\zeta(\textbf{r}) = (\exp(-(y-y_1)/d_\mathrm{step})+1)^{-1} + (\exp(-(y_2-y)/d_\mathrm{step})+1)^{-1} $, $y_1 = 1607$ nm, $y_2 = -49$ nm, and $d_\mathrm{step} = 2$ nm. In the rest of the system, $\textbf{A} = (-yB, 0)$ is used. The resulting $A_x$ and $A_y$ profiles are shown in \autoref{fig:fig6}.

\paragraph*{Hofstadter butterfly calculation}

For the calculation of Hofstadter butterfly one has to consider a magnetic unit cell whose
length is the least common multiple of the lattice periodicity and the periodicity introduced by the Peierls phase. For graphene, it contains more than hundreds of thousands of carbon atoms when the magnetic field strength is smaller than 1 T. However, the calculation is greatly simplified by considering a graphene ribbon. %That is the approach adopted in this work. 
For an armchair ribbon with translational invariance along the $x$ direction and finite width along the $y$ direction, in the presence of a perpendicular magnetic field, dispersionless Landau levels appear near $k_{x}=0$, and the dispersive edge states show up at larger $k_{x}$. 
Calculating $E_{k_{x}=0}$ as a function of magnetic field, we get the Hofstadter butterfly of graphene. Because of the finite width of the ribbon, the spectrum can also contain edge states. With the increase of $B$ the Landau levels elongate, and at some $B$ the edge states are pushed to $k_x=0$, which results in the appearance of the states in the gaps. 
To lower the computational burden, we use the scalable tight-binding model \cite{Liu2015} with $s_\mathrm{f}\sim 8$ to calculate $E_{k_{x}=0}$ as a function of magnetic field for an armchair ribbon with periodic length along the $x$ axis equal to the superlattice period ($\lambda$). Here, in order to ensure superlattice period equal to a multiple of $3a$,  $s_\mathrm{f}$ is not an integer, and the ribbon width is larger than $20 \lambda$ to show the superlattice effect. 

In transport, only the states at the Fermi level contribute to the conductance. Therefore, we calculate Hofstadter butterfly spectra for all $V_\mathrm{tg}$ values and collect the Fermi states to construct the gate-dependent Hofstadter butterfly spectrum to compare with $R_{xx}$ and $G_\mathrm{H}$
obtained from the transport calculation.
 Note that the spectrum in \autoref{fig:fig5}(g) contains energy levels that appear across the gaps, which are an artifact of the method due to the finite width of the ribbon. As mentioned above, they appear since at some value of $B$ the edge states are pushed to $k_x=0$. 
\\

%
%\bmhead{Data availability} Additional data related to this paper may be requested from the authors.
%
%
%\bmhead{Code availability} Code may be requested from the authors upon reasonable request.

%\begin{acknowledgments}
\bmhead{Acknowledgments}
We thank National Science and Technology Council of Taiwan (grant numbers: MOST 109-2112-M-006-020-MY3 and NSTC 112-2112-M-006-019-MY3) for financial supports and National Center for High-performance Computing (NCHC) for providing computational and storage resources. This
research was supported in part by PL-Grid Infrastructure, and by the program ,,Excellence Initiative -- Research University'' for the AGH University of Science and Technology.
%\end{acknowledgments}

%\bmhead{Authors contributions}
%A.M.K. and M.-H.L. performed transport calculations and wrote the manuscript with the input from all authors. 
%S.-C.C. calculated the mini-band structures within the continuum model and the Hofstadter butterfly within the tight-binding model. M.-H.L. guided the project. 
%%
%
%\bmhead{Competing interests} The Authors declare no Competing Financial or Non-Financial Interests.

%
%\bibliographystyle{naturemag_noURL}
%\bibliography{SL}

\begin{thebibliography}{10}
\expandafter\ifx\csname url\endcsname\relax
  \def\url#1{\texttt{#1}}\fi
\expandafter\ifx\csname urlprefix\endcsname\relax\def\urlprefix{URL }\fi
\providecommand{\bibinfo}[2]{#2}
\providecommand{\eprint}[2][]{\url{#2}}

\bibitem{Wallbank2013}
\bibinfo{author}{Wallbank, J.~R.}, \bibinfo{author}{Patel, A.~A.},
  \bibinfo{author}{Mucha-Kruczy\ifmmode~\acute{n}\else \'{n}\fi{}ski, M.},
  \bibinfo{author}{Geim, A.~K.} \& \bibinfo{author}{Fal'ko, V.~I.}
\newblock \bibinfo{title}{{Generic miniband structure of graphene on a
  hexagonal substrate}}.
\newblock \emph{\bibinfo{journal}{Phys. Rev. B}} \textbf{\bibinfo{volume}{87}},
  \bibinfo{pages}{245408} (\bibinfo{year}{2013}).

\bibitem{Park2008}
\bibinfo{author}{Park, C.-H.}, \bibinfo{author}{Yang, L.},
  \bibinfo{author}{Son, Y.-W.}, \bibinfo{author}{Cohen, M.~L.} \&
  \bibinfo{author}{Louie, S.~G.}
\newblock \bibinfo{title}{{Anisotropic behaviours of massless Dirac fermions in
  graphene under periodic potentials}}.
\newblock \emph{\bibinfo{journal}{Nat. Phys.}} \textbf{\bibinfo{volume}{4}},
  \bibinfo{pages}{213--217} (\bibinfo{year}{2008}).

\bibitem{Park2008prb}
\bibinfo{author}{Park, C.-H.}, \bibinfo{author}{Yang, L.},
  \bibinfo{author}{Son, Y.-W.}, \bibinfo{author}{Cohen, M.~L.} \&
  \bibinfo{author}{Louie, S.~G.}
\newblock \bibinfo{title}{{New Generation of Massless Dirac Fermions in
  Graphene under External Periodic Potentials}}.
\newblock \emph{\bibinfo{journal}{Phys. Rev. Lett.}}
  \textbf{\bibinfo{volume}{101}}, \bibinfo{pages}{126804}
  (\bibinfo{year}{2008}).

\bibitem{Brey2009}
\bibinfo{author}{Brey, L.} \& \bibinfo{author}{Fertig, H.~A.}
\newblock \bibinfo{title}{{Emerging Zero Modes for Graphene in a Periodic
  Potential}}.
\newblock \emph{\bibinfo{journal}{Phys. Rev. Lett.}}
  \textbf{\bibinfo{volume}{103}}, \bibinfo{pages}{046809}
  (\bibinfo{year}{2009}).

\bibitem{Barbier2010}
\bibinfo{author}{Barbier, M.}, \bibinfo{author}{Vasilopoulos, P.} \&
  \bibinfo{author}{Peeters, F.~M.}
\newblock \bibinfo{title}{{Extra Dirac points in the energy spectrum for
  superlattices on single-layer graphene}}.
\newblock \emph{\bibinfo{journal}{Phys. Rev. B}} \textbf{\bibinfo{volume}{81}},
  \bibinfo{pages}{075438} (\bibinfo{year}{2010}).

\bibitem{Kang2020}
\bibinfo{author}{Kang, W.-H.}, \bibinfo{author}{Chen, S.-C.} \&
  \bibinfo{author}{Liu, M.-H.}
\newblock \bibinfo{title}{Cloning of zero modes in one-dimensional graphene
  superlattices}.
\newblock \emph{\bibinfo{journal}{Phys. Rev. B}}
  \textbf{\bibinfo{volume}{102}}, \bibinfo{pages}{195432}
  (\bibinfo{year}{2020}).

\bibitem{Sun2011}
\bibinfo{author}{Sun, Z.} \emph{et~al.}
\newblock \bibinfo{title}{{Towards hybrid superlattices in graphene}}.
\newblock \emph{\bibinfo{journal}{Nat. Commun.}} \textbf{\bibinfo{volume}{2}},
  \bibinfo{pages}{559} (\bibinfo{year}{2011}).

\bibitem{Zhang2018}
\bibinfo{author}{Zhang, Y.}, \bibinfo{author}{Kim, Y.},
  \bibinfo{author}{Gilbert, M.~J.} \& \bibinfo{author}{Mason, N.}
\newblock \bibinfo{title}{{Electronic transport in a two-dimensional
  superlattice engineered via self-assembled nanostructures}}.
\newblock \emph{\bibinfo{journal}{npj 2D Mater. Appl.}}
  \textbf{\bibinfo{volume}{2}}, \bibinfo{pages}{31} (\bibinfo{year}{2018}).

\bibitem{Xue2011}
\bibinfo{author}{Xue, J.} \emph{et~al.}
\newblock \bibinfo{title}{{Scanning tunnelling microscopy and spectroscopy of
  ultra-flat graphene on hexagonal boron nitride}}.
\newblock \emph{\bibinfo{journal}{Nat. Mater.}} \textbf{\bibinfo{volume}{10}},
  \bibinfo{pages}{282--285} (\bibinfo{year}{2011}).

\bibitem{Decker2011}
\bibinfo{author}{Decker, R.} \emph{et~al.}
\newblock \bibinfo{title}{{Local Electronic Properties of Graphene on a BN
  Substrate via Scanning Tunneling Microscopy}}.
\newblock \emph{\bibinfo{journal}{Nano Lett.}} \textbf{\bibinfo{volume}{11}},
  \bibinfo{pages}{2291--2295} (\bibinfo{year}{2011}).

\bibitem{Yankowitz2012}
\bibinfo{author}{Yankowitz, M.} \emph{et~al.}
\newblock \bibinfo{title}{{Emergence of superlattice Dirac points in graphene
  on hexagonal boron nitride}}.
\newblock \emph{\bibinfo{journal}{Nat. Phys.}} \textbf{\bibinfo{volume}{8}},
  \bibinfo{pages}{382--386} (\bibinfo{year}{2012}).

\bibitem{Ponomarenko2013}
\bibinfo{author}{Ponomarenko, L.~A.} \emph{et~al.}
\newblock \bibinfo{title}{{Cloning of Dirac fermions in graphene
  superlattices}}.
\newblock \emph{\bibinfo{journal}{Nature}} \textbf{\bibinfo{volume}{497}},
  \bibinfo{pages}{594--597} (\bibinfo{year}{2013}).

\bibitem{Hunt2013}
\bibinfo{author}{Hunt, B.} \emph{et~al.}
\newblock \bibinfo{title}{{Massive Dirac Fermions and Hofstadter Butterfly in a
  van der Waals Heterostructure}}.
\newblock \emph{\bibinfo{journal}{Science}} \textbf{\bibinfo{volume}{340}},
  \bibinfo{pages}{1427--1430} (\bibinfo{year}{2013}).

\bibitem{Dean2013}
\bibinfo{author}{Dean, C.~R.} \emph{et~al.}
\newblock \bibinfo{title}{{Hofstadter's butterfly and the fractal quantum Hall
  effect in moir{\'e} superlattices}}.
\newblock \emph{\bibinfo{journal}{Nature}} \textbf{\bibinfo{volume}{497}},
  \bibinfo{pages}{598--602} (\bibinfo{year}{2013}).

\bibitem{Yu2014}
\bibinfo{author}{Yu, G.~L.} \emph{et~al.}
\newblock \bibinfo{title}{{Hierarchy of Hofstadter states and replica quantum
  Hall ferromagnetism in graphene superlattices}}.
\newblock \emph{\bibinfo{journal}{Nat. Phys.}} \textbf{\bibinfo{volume}{10}},
  \bibinfo{pages}{525--529} (\bibinfo{year}{2014}).

\bibitem{Burg2019}
\bibinfo{author}{Burg, G.~W.} \emph{et~al.}
\newblock \bibinfo{title}{{Correlated Insulating States in Twisted Double
  Bilayer Graphene}}.
\newblock \emph{\bibinfo{journal}{Phys. Rev. Lett.}}
  \textbf{\bibinfo{volume}{123}}, \bibinfo{pages}{197702}
  (\bibinfo{year}{2019}).

\bibitem{Shen2020}
\bibinfo{author}{Shen, C.} \emph{et~al.}
\newblock \bibinfo{title}{Correlated states in twisted double bilayer
  graphene}.
\newblock \emph{\bibinfo{journal}{Nat. Phys.}} \textbf{\bibinfo{volume}{16}},
  \bibinfo{pages}{520--525} (\bibinfo{year}{2020}).

\bibitem{Liu2020}
\bibinfo{author}{Liu, X.} \emph{et~al.}
\newblock \bibinfo{title}{Tunable spin-polarized correlated states in twisted
  double bilayer graphene}.
\newblock \emph{\bibinfo{journal}{Nature}} \textbf{\bibinfo{volume}{583}},
  \bibinfo{pages}{221--225} (\bibinfo{year}{2020}).

\bibitem{Lin2020}
\bibinfo{author}{Lin, F.} \emph{et~al.}
\newblock \bibinfo{title}{{Heteromoir{\'e} Engineering on Magnetic Bloch
  Transport in Twisted Graphene Superlattices}}.
\newblock \emph{\bibinfo{journal}{Nano Lett.}} \textbf{\bibinfo{volume}{20}},
  \bibinfo{pages}{7572--7579} (\bibinfo{year}{2020}).

\bibitem{deVries2020}
\bibinfo{author}{de~Vries, F.~K.} \emph{et~al.}
\newblock \bibinfo{title}{{Combined Minivalley and Layer Control in Twisted
  Double Bilayer Graphene}}.
\newblock \emph{\bibinfo{journal}{Phys. Rev. Lett.}}
  \textbf{\bibinfo{volume}{125}}, \bibinfo{pages}{176801}
  (\bibinfo{year}{2020}).

\bibitem{Rickhaus2021}
\bibinfo{author}{Rickhaus, P.} \emph{et~al.}
\newblock \bibinfo{title}{Correlated electron-hole state in twisted
  double-bilayer graphene}.
\newblock \emph{\bibinfo{journal}{Science}} \textbf{\bibinfo{volume}{373}},
  \bibinfo{pages}{1257--1260} (\bibinfo{year}{2021}).

\bibitem{WangLujun2019}
\bibinfo{author}{Wang, L.} \emph{et~al.}
\newblock \bibinfo{title}{{New Generation of Moir{\'e} Superlattices in Doubly
  Aligned hBN/Graphene/hBN Heterostructures}}.
\newblock \emph{\bibinfo{journal}{Nano Lett.}} \textbf{\bibinfo{volume}{19}},
  \bibinfo{pages}{2371--2376} (\bibinfo{year}{2019}).

\bibitem{WangZihao2019}
\bibinfo{author}{Wang, Z.} \emph{et~al.}
\newblock \bibinfo{title}{{Composite super-moir{\'e} lattices in double-aligned
  graphene heterostructures}}.
\newblock \emph{\bibinfo{journal}{Sci. Adv.}} \textbf{\bibinfo{volume}{5}},
  \bibinfo{pages}{eaay8897} (\bibinfo{year}{2019}).

\bibitem{Kumar2017}
\bibinfo{author}{Kumar, R.~K.} \emph{et~al.}
\newblock \bibinfo{title}{{High-temperature quantum oscillations caused by
  recurring Bloch states in graphene superlattices}}.
\newblock \emph{\bibinfo{journal}{Science}} \textbf{\bibinfo{volume}{357}},
  \bibinfo{pages}{181--184} (\bibinfo{year}{2017}).

\bibitem{Barrier2020}
\bibinfo{author}{Barrier, J.} \emph{et~al.}
\newblock \bibinfo{title}{{Long-range ballistic transport of Brown-Zak fermions
  in graphene superlattices}}.
\newblock \emph{\bibinfo{journal}{Nat. Commun.}} \textbf{\bibinfo{volume}{11}},
  \bibinfo{pages}{5756} (\bibinfo{year}{2020}).

\bibitem{Wang2015}
\bibinfo{author}{Wang, L.} \emph{et~al.}
\newblock \bibinfo{title}{{Evidence for a fractional fractal quantum Hall
  effect in graphene superlattices}}.
\newblock \emph{\bibinfo{journal}{Science}} \textbf{\bibinfo{volume}{350}},
  \bibinfo{pages}{1231--1234} (\bibinfo{year}{2015}).

\bibitem{Andrews2020}
\bibinfo{author}{Andrews, B.} \& \bibinfo{author}{Soluyanov, A.}
\newblock \bibinfo{title}{{Fractional quantum Hall states for moir\'e
  superstructures in the Hofstadter regime}}.
\newblock \emph{\bibinfo{journal}{Phys. Rev. B}}
  \textbf{\bibinfo{volume}{101}}, \bibinfo{pages}{235312}
  (\bibinfo{year}{2020}).

\bibitem{Dubey2013}
\bibinfo{author}{Dubey, S.} \emph{et~al.}
\newblock \bibinfo{title}{{Tunable Superlattice in Graphene To Control the
  Number of Dirac Points}}.
\newblock \emph{\bibinfo{journal}{Nano Lett.}} \textbf{\bibinfo{volume}{13}},
  \bibinfo{pages}{3990--3995} (\bibinfo{year}{2013}).

\bibitem{Drienovsky2014}
\bibinfo{author}{Drienovsky, M.} \emph{et~al.}
\newblock \bibinfo{title}{{Towards superlattices: Lateral bipolar multibarriers
  in graphene}}.
\newblock \emph{\bibinfo{journal}{Phys. Rev. B}} \textbf{\bibinfo{volume}{89}},
  \bibinfo{pages}{115421} (\bibinfo{year}{2014}).

\bibitem{Kuiri2018}
\bibinfo{author}{Kuiri, M.}, \bibinfo{author}{Gupta, G.~K.},
  \bibinfo{author}{Ronen, Y.}, \bibinfo{author}{Das, T.} \&
  \bibinfo{author}{Das, A.}
\newblock \bibinfo{title}{{Large Landau-level splitting in a tunable
  one-dimensional graphene superlattice probed by magnetocapacitance
  measurements}}.
\newblock \emph{\bibinfo{journal}{Phys. Rev. B}} \textbf{\bibinfo{volume}{98}},
  \bibinfo{pages}{035418} (\bibinfo{year}{2018}).

\bibitem{Forsythe2018}
\bibinfo{author}{Forsythe, C.} \emph{et~al.}
\newblock \bibinfo{title}{{Band structure engineering of 2D materials using
  patterned dielectric superlattices}}.
\newblock \emph{\bibinfo{journal}{Nat. Nanotechnol.}}
  \textbf{\bibinfo{volume}{13}}, \bibinfo{pages}{566--571}
  (\bibinfo{year}{2018}).

\bibitem{Li2021}
\bibinfo{author}{Li, Y.} \emph{et~al.}
\newblock \bibinfo{title}{Anisotropic band flattening in graphene with
  one-dimensional superlattices}.
\newblock \emph{\bibinfo{journal}{Nat. Nanotechnol.}}
  \textbf{\bibinfo{volume}{16}}, \bibinfo{pages}{525--530}
  (\bibinfo{year}{2021}).

\bibitem{Drienovsky2018}
\bibinfo{author}{Drienovsky, M.} \emph{et~al.}
\newblock \bibinfo{title}{{Commensurability Oscillations in One-Dimensional
  Graphene Superlattices}}.
\newblock \emph{\bibinfo{journal}{Phys. Rev. Lett.}}
  \textbf{\bibinfo{volume}{121}}, \bibinfo{pages}{026806}
  (\bibinfo{year}{2018}).

\bibitem{Ruiz2022}
\bibinfo{author}{Barcons~Ruiz, D.} \emph{et~al.}
\newblock \bibinfo{title}{Engineering high quality graphene superlattices via
  ion milled ultra-thin etching masks}.
\newblock \emph{\bibinfo{journal}{Nat. Commun.}} \textbf{\bibinfo{volume}{13}},
  \bibinfo{pages}{6926} (\bibinfo{year}{2022}).

\bibitem{Huber2020}
\bibinfo{author}{Huber, R.} \emph{et~al.}
\newblock \bibinfo{title}{{Gate-Tunable Two-Dimensional Superlattices in
  Graphene}}.
\newblock \emph{\bibinfo{journal}{Nano Lett.}} \textbf{\bibinfo{volume}{20}},
  \bibinfo{pages}{8046--8052} (\bibinfo{year}{2020}).

\bibitem{Huber2022}
\bibinfo{author}{Huber, R.} \emph{et~al.}
\newblock \bibinfo{title}{Band conductivity oscillations in a gate-tunable
  graphene superlattice}.
\newblock \emph{\bibinfo{journal}{Nat. Commun.}} \textbf{\bibinfo{volume}{13}},
  \bibinfo{pages}{2856} (\bibinfo{year}{2022}).

\bibitem{Taychatanapat2013}
\bibinfo{author}{Taychatanapat, T.}, \bibinfo{author}{Watanabe, K.},
  \bibinfo{author}{Taniguchi, T.} \& \bibinfo{author}{Jarillo-Herrero, P.}
\newblock \bibinfo{title}{Electrically tunable transverse magnetic focusing in
  graphene}.
\newblock \emph{\bibinfo{journal}{Nat. Phys.}} \textbf{\bibinfo{volume}{9}},
  \bibinfo{pages}{225--229} (\bibinfo{year}{2013}).

\bibitem{Rao2023}
\bibinfo{author}{Rao, Q.} \emph{et~al.}
\newblock \bibinfo{title}{{Ballistic transport spectroscopy of
  spin-orbit-coupled bands in monolayer graphene on WSe$_2$}}.
\newblock \bibinfo{howpublished}{Preprint at
  \url{https://arxiv.org/abs/2303.01018}} (\bibinfo{year}{2023}).

\bibitem{Lee2016}
\bibinfo{author}{Lee, M.} \emph{et~al.}
\newblock \bibinfo{title}{{Ballistic miniband conduction in a graphene
  superlattice}}.
\newblock \emph{\bibinfo{journal}{Science}} \textbf{\bibinfo{volume}{353}},
  \bibinfo{pages}{1526--1529} (\bibinfo{year}{2016}).

\bibitem{Berdyugin2020}
\bibinfo{author}{Berdyugin, A.~I.} \emph{et~al.}
\newblock \bibinfo{title}{{Minibands in twisted bilayer graphene probed by
  magnetic focusing}}.
\newblock \emph{\bibinfo{journal}{Sci. Adv.}} \textbf{\bibinfo{volume}{6}},
  \bibinfo{pages}{eaay7838} (\bibinfo{year}{2020}).

\bibitem{Milovanovic2014}
\bibinfo{author}{Milovanovi\'c, S.~P.}, \bibinfo{author}{Ramezani~Masir, M.} \&
  \bibinfo{author}{Peeters, F.~M.}
\newblock \bibinfo{title}{Magnetic electron focusing and tuning of the electron
  current with a pn-junction}.
\newblock \emph{\bibinfo{journal}{J. Appl. Phys.}}
  \textbf{\bibinfo{volume}{115}}, \bibinfo{pages}{043719}
  (\bibinfo{year}{2014}).

\bibitem{Chen2016}
\bibinfo{author}{Chen, S.} \emph{et~al.}
\newblock \bibinfo{title}{Electron optics with p-n junctions in ballistic
  graphene}.
\newblock \emph{\bibinfo{journal}{Science}} \textbf{\bibinfo{volume}{353}},
  \bibinfo{pages}{1522--1525} (\bibinfo{year}{2016}).

\bibitem{Chen2020}
\bibinfo{author}{Chen, S.-C.}, \bibinfo{author}{Kraft, R.},
  \bibinfo{author}{Danneau, R.}, \bibinfo{author}{Richter, K.} \&
  \bibinfo{author}{Liu, M.-H.}
\newblock \bibinfo{title}{Electrostatic superlattices on scaled graphene
  lattices}.
\newblock \emph{\bibinfo{journal}{Commun. Phys.}} \textbf{\bibinfo{volume}{3}},
  \bibinfo{pages}{71} (\bibinfo{year}{2020}).

\bibitem{Kraft2020}
\bibinfo{author}{Kraft, R.} \emph{et~al.}
\newblock \bibinfo{title}{{Anomalous Cyclotron Motion in Graphene Superlattice
  Cavities}}.
\newblock \emph{\bibinfo{journal}{Phys. Rev. Lett.}}
  \textbf{\bibinfo{volume}{125}}, \bibinfo{pages}{217701}
  (\bibinfo{year}{2020}).

\bibitem{Ashcroft1976}
\bibinfo{author}{Ashcroft, N.~W.} \& \bibinfo{author}{Mermin, N.~D.}
\newblock \emph{\bibinfo{title}{{Solid State Physics}}} (\bibinfo{year}{1976}).

\bibitem{Stegmann2015}
\bibinfo{author}{Stegmann, T.} \& \bibinfo{author}{Lorke, A.}
\newblock \bibinfo{title}{{Edge magnetotransport in graphene: A combined
  analytical and numerical study}}.
\newblock \emph{\bibinfo{journal}{Ann. Phys.}} \textbf{\bibinfo{volume}{527}},
  \bibinfo{pages}{723--736} (\bibinfo{year}{2015}).

\bibitem{Petrovic2017}
\bibinfo{author}{Petrovi{\'{c}}, M.~D.}, \bibinfo{author}{Milovanovi{\'{c}},
  S.~P.} \& \bibinfo{author}{Peeters, F.~M.}
\newblock \bibinfo{title}{Scanning gate microscopy of magnetic focusing in
  graphene devices: quantum versus classical simulation}.
\newblock \emph{\bibinfo{journal}{Nanotechnology}}
  \textbf{\bibinfo{volume}{28}}, \bibinfo{pages}{185202}
  (\bibinfo{year}{2017}).

\bibitem{Buttiker1986}
\bibinfo{author}{B\"uttiker, M.}
\newblock \bibinfo{title}{{Four-Terminal Phase-Coherent Conductance}}.
\newblock \emph{\bibinfo{journal}{Phys. Rev. Lett.}}
  \textbf{\bibinfo{volume}{57}}, \bibinfo{pages}{1761--1764}
  (\bibinfo{year}{1986}).

\bibitem{Datta1995}
\bibinfo{author}{Datta, S.}
\newblock \emph{\bibinfo{title}{{Electronic Transport in Mesoscopic Systems}}}
  (\bibinfo{publisher}{Cambridge University Press, Cambridge},
  \bibinfo{year}{1995}).

\bibitem{Thouless1982}
\bibinfo{author}{Thouless, D.~J.}, \bibinfo{author}{Kohmoto, M.},
  \bibinfo{author}{Nightingale, M.~P.} \& \bibinfo{author}{den Nijs, M.}
\newblock \bibinfo{title}{{Quantized Hall Conductance in a Two-Dimensional
  Periodic Potential}}.
\newblock \emph{\bibinfo{journal}{Phys. Rev. Lett.}}
  \textbf{\bibinfo{volume}{49}}, \bibinfo{pages}{405--408}
  (\bibinfo{year}{1982}).

\bibitem{Streda1982}
\bibinfo{author}{Streda, P.}
\newblock \bibinfo{title}{{Quantised Hall effect in a two-dimensional periodic
  potential}}.
\newblock \emph{\bibinfo{journal}{J. Phys. C: Solid State Phys.}}
  \textbf{\bibinfo{volume}{15}}, \bibinfo{pages}{L1299} (\bibinfo{year}{1982}).

\bibitem{deVries2021}
\bibinfo{author}{de~Vries, F.~K.} \emph{et~al.}
\newblock \bibinfo{title}{{Gate-defined Josephson junctions in magic-angle
  twisted bilayer graphene}}.
\newblock \emph{\bibinfo{journal}{Nat. Nanotechnol.}}
  \textbf{\bibinfo{volume}{16}}, \bibinfo{pages}{760--763}
  (\bibinfo{year}{2021}).

\bibitem{Krix2022}
\bibinfo{author}{Krix, Z.~E.} \& \bibinfo{author}{Sushkov, O.~P.}
\newblock \bibinfo{title}{Patterned bilayer graphene as a tunable strongly
  correlated system}.
\newblock \emph{\bibinfo{journal}{Phys. Rev. B}}
  \textbf{\bibinfo{volume}{107}}, \bibinfo{pages}{165158}
  (\bibinfo{year}{2023}).

\bibitem{Liu2015}
\bibinfo{author}{Liu, M.-H.} \emph{et~al.}
\newblock \bibinfo{title}{{Scalable Tight-Binding Model for Graphene}}.
\newblock \emph{\bibinfo{journal}{Phys. Rev. Lett.}}
  \textbf{\bibinfo{volume}{114}}, \bibinfo{pages}{036601}
  (\bibinfo{year}{2015}).

\bibitem{Baranger1989}
\bibinfo{author}{Baranger, H.~U.} \& \bibinfo{author}{Stone, A.~D.}
\newblock \bibinfo{title}{{Electrical linear-response theory in an arbitrary
  magnetic field: A new Fermi-surface formation}}.
\newblock \emph{\bibinfo{journal}{Phys. Rev. B}} \textbf{\bibinfo{volume}{40}},
  \bibinfo{pages}{8169--8193} (\bibinfo{year}{1989}).
\end{thebibliography}

\end{document}